\pdfoutput=1
\documentclass[10pt,english,journal]{IEEEtran}
\usepackage[T1]{fontenc}
\usepackage[latin9]{inputenc}
\usepackage{array}
\usepackage{pmboxdraw}
\usepackage{multirow}
\usepackage{amsmath}
\usepackage{amssymb}
\usepackage{stackrel}
\usepackage{graphicx}

\makeatletter

\providecommand{\tabularnewline}{\\}

\usepackage{xcolor,soul}
\sethlcolor{lightgray}

\newcommand{\hilight}[1]{\colorbox{lightgray}{#1}}

\usepackage{amsmath}

\makeatother

\usepackage{babel}
\begin{document}

\title{Cross-Link Interference Suppression By Orthogonal Projector For 5G
Dynamic TDD URLLC Systems}

\author{\IEEEauthorblockN{Ali A. Esswie$^{1,2}$,\textit{ Member, IEEE}, and\textit{ }Klaus
I. Pedersen\textit{$^{1,2}$, Senior Member, IEEE}\\
$^{1}$Nokia Bell-Labs, Aalborg, Denmark\\
$^{2}$Department of Electronic Systems, Aalborg University, Denmark}}
\maketitle
\begin{abstract}
Dynamic time division duplexing (TDD) is envisioned as a vital transmission
technology of the 5G new radio, due to its reciprocal propagation
characteristics. However, the potential cross-link interference (CLI)
imposes a fundamental limitation against the feasibility of the ultra-reliable
and low latency communications (URLLC) in dynamic-TDD systems. In
this work, we propose a near-optimal and complexity-efficient CLI
suppression scheme using orthogonal spatial projection, while the
signaling overhead is limited to $\textnormal{B-bit}$, over the back-haul
links. Compared to the state-of-the-art dynamic-TDD studies, proposed
solution offers a significant improvement of the URLLC outage latency,
e.g., $\sim-199\%$ reduction, while boosting the achievable capacity
per the URLLC packet by $\sim+156\%$. 

\end{abstract}



\IEEEpeerreviewmaketitle{}

\section{Introduction}

\IEEEPARstart{U}{}ltra-reliable and low-latency communication (URLLC)
is the major service class of the 5G new radio (NR) {[}1{]}. URLLC
denotes short and stochastic packet transmissions with extreme reliability
and radio latency bounds, i.e., couple of milli-seconds with a success
probability of $99.999\%$ {[}2{]}. Furthermore, the global regulatory
bodies have envisioned early 5G deployments over the 3.5 GHz spectrum
due to its abundant available unpaired bands. Accordingly, dynamic
time division duplexing (TDD) has become of a great significance {[}3{]}.
With dynamic TDD, base-stations (BSs) independently and dynamically
in time select their respective link directions based on individual
objective functions, leading to an improved transmission adaptation
to the sporadic traffic arrivals.

However, the URLLC reliability and latency targets are further challenging
to achieve in dynamic TDD systems {[}2{]} due to: (a) the switching
time between the downlink (DL) and uplink (UL) sub-frames, and (b)
the potential inter-cell cross-link interference (CLI) between neighboring
BSs of different directional transmissions {[}4{]}. That is, the DL-to-UL
CLI (BS-BS) and UL-to-DL CLI (user-equipment to user-equipment (UE-UE)).
The former is tackled by the flexible frame design of the 5G-NR, where
variable transmission time intervals (TTIs) and a scalable sub-carrier
spacing (SCS) are supported {[}1{]}. Thus, the DL and UL switching
delay can be slot-dependent, i.e., $\ll1$ ms. Although, the latter
issue, especially the BS-BS CLI due to the power imbalance between
the DL and UL transmissions, remains a critical issue against practical
implementation of the dynamic TDD macro systems. 

As part of the long-term evolution, i.e., 4G, standards, advanced
linear interference rejection combining (IRC) transceivers {[}5{]}
are adopted to suppress the inter-cell interference sub-space from
that is of the useful signal. Although, within dense macro deployments,
there exist multiple dominant and sparse BS-BS CLI interferers, degrading
the IRC decoding performance due to the linear interference averaging.
Accordingly, optimal BS-BS CLI cancellation {[}6{]} is discussed within
3GPP, where inter-cell full-packet exchange is assumed. Moreover,
coordinated dynamic scheduling and beam-forming {[}7, 8{]} are proposed
to counteract the CLI by globalizing the BS scheduling decisions.
Furthermore, joint beam-forming schemes are suggested {[}9, 10{]}
in order to control the inflicted inter-cell CLI in the spatial domain.
On another side, opportunistic CLI pre-avoidance {[}4, 11, 12{]} schemes
have been introduced based on ordered signal-to-interference-noise-ratio
(SINR) lists and a sliding radio frame configuration (RFC) code-book
design, respectively.

In this paper, we propose a high-performance and low-complexity BS-BS
CLI suppression algorithm (CSA) for 5G-NR dynamic TDD macro systems.
The proposed scheme utilizes a linear estimation of an orthonormal
sub-space projector to reliably suppress the BS-BS CLI on-the-fly,
while it combines a hybrid radio frame design, cyclic-offset based
frame code-book, and dual-objective dynamic user scheduling to opportunistically
pre-avoid the UE-UE CLI occurrence. Compared to state-of-the-art dynamic-TDD
studies, the proposed scheme offers a significant enhancement of the
URLLC UL and DL outage latency, while improving the ergodic capacity,
approaching the optimal CLI-free case. However, the proposed scheme
neither requires periodic user CLI measurements nor significant signaling
overhead. Particularly, the contribution aspects of this paper are
as follows:
\begin{itemize}
\item Unlike the standard linear IRC receiver, we utilize a newly proposed
inter-BS exchange of the user DL spatial signatures to manipulate
the estimated interference covariance. Hence, we introduce an enhanced
formulation of the standard IRC receiver, where the BS-BS CLI spatial
span is regularized \textit{on-the-fly}, leading the IRC receiver
be further directive to the user effective channel.
\item The proposed solution requires a modest inter-BS signaling overhead. 
\item The proposed enhanced IRC receiver provides $\sim199\%$ gain of the
achievable URLLC outage latency, compared to state-of-the-art relevant
IRC literature.
\end{itemize}
Due to the complexity of the addressed problem herein and the 5G-NR
system dynamics, the performance of the proposed solution is assessed
using a highly-detailed system level simulator, with a high degree
of realism. Following the same simulation methodology in {[}4{]},
these simulations are based on widely-accepted mathematical models
and being validated against the latest 3GPP 5G-NR assumptions. The
main functionalities of Layer 1 and 2 of the 5G-NR protocol stack
are integrated including the hybrid automatic repeat request (HARQ)
re-transmissions, 3D spatial channel modeling, adaptive modulation
and coding.

The paper is organized as follows. Section II introduces the system
model of this work. Section III presents the proposed solution while
Section IV discusses the performance assessment metrics. Conclusions
are drawn in Section V. 

\section{System Model}

We consider a synchronous dynamic-TDD 5G-NR network of a single cluster
of $C$ BSs, each equipped with $N$ antennas. There are $K^{\textnormal{dl}}$
and $K^{\textnormal{ul}}$ uniformly-distributed DL and UL active
UEs per BS, respectively, each with $M$ antennas. The URLLC stochastic
FTP3 traffic model is assumed, with finite payload sizes of $\textnormal{\ensuremath{\mathit{f}^{dl}}}$
and $\textnormal{\ensuremath{\mathit{f}^{ul}}}$ bits, and Poisson
arrival processes $\textnormal{\ensuremath{\lambda}}^{\textnormal{dl}}$
and $\textnormal{\ensuremath{\lambda}}^{\textnormal{ul}},$ in the
DL and UL directions. Hence, the directional offered loads per BS
are given by: $\varOmega^{\textnormal{\{dl,ul\}}}=K^{\textnormal{\{dl, ul\}}}\times\textnormal{\ensuremath{\mathit{f}^{\textnormal{\{dl,ul\}}}}}\times\textnormal{\ensuremath{\lambda}}^{\textnormal{\{dl,ul\}}}$,
with $\varOmega=\varOmega^{\textnormal{dl}}+\varOmega^{\textnormal{ul}}$
as the total load per cell. 

We adopt the latest system assumptions of the 3GPP specifications
for URLLC {[}2{]}. Hence, a 10-ms RFC is composed of 10 sub-frames,
each is constructed of a scalable number of slots. Accordingly, we
consider the dynamic 3GPP release-15 slot format design {[}13{]},
with a flexible structure of the DL, UL and special symbols, respectively,
as shown in Fig. 1. UEs are dynamically multiplexed by the orthogonal
frequency division multiple access (OFDMA), with 30 KHz SCS and a
physical resource block (PRB) of 12 consecutive SCs. Furthermore,
a short TTI duration of 4-OFDM symbols is adopted. 

\begin{figure}
\begin{centering}
\includegraphics[scale=0.8]{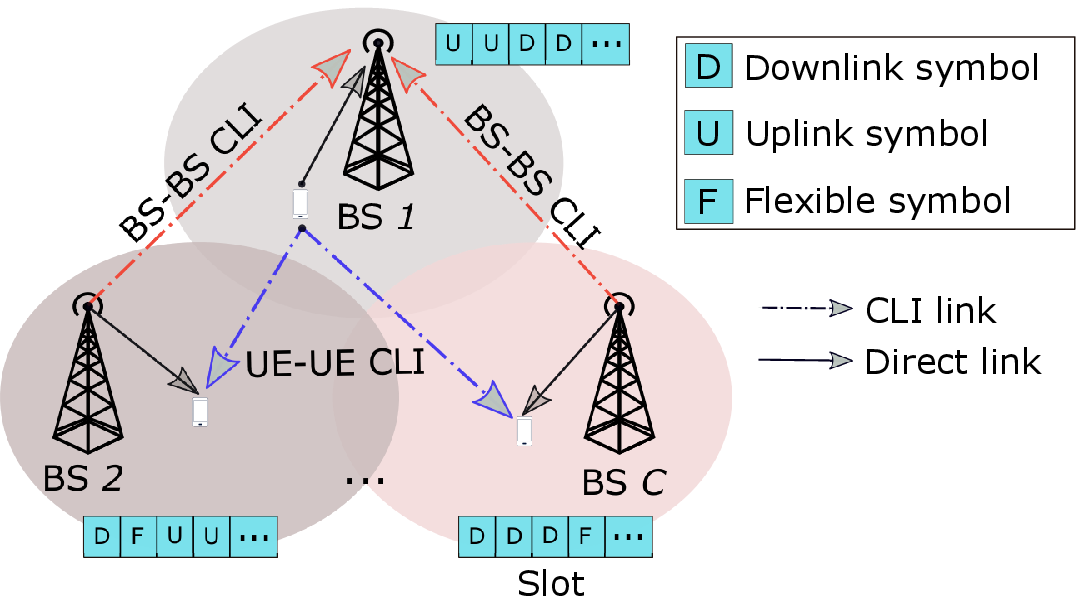}
\par\end{centering}
\centering{}{\small{}Fig. 1. Flexible TDD system modeling.}{\small \par}
\end{figure}

Consider $\text{\ensuremath{\mathfrak{B}}}_{\textnormal{dl}},$ $\text{\ensuremath{\mathfrak{B}}}_{\textnormal{ul}}$,
$\text{\ensuremath{\mathcal{K}}}_{\textnormal{dl}}$ and $\text{\ensuremath{\mathcal{K}}}_{\textnormal{ul}}$
as the sets of BSs and UEs with DL and UL transmissions, respectively.
Thus, the DL signal at the $k^{th}$ UE, where $k\text{\ensuremath{\in\text{\ensuremath{\mathcal{K}}}_{\textnormal{dl}}}}$,
$c_{k}\text{\ensuremath{\in\text{\ensuremath{\mathfrak{B}}}_{\textnormal{dl}}}}$,
is expressed by

{\small{}
\begin{equation}
\boldsymbol{\textnormal{y}}_{k,c_{k}}^{\textnormal{dl}}=\underbrace{\boldsymbol{\textnormal{\textbf{H}}}_{k,c_{k}}^{\textnormal{dl}}\boldsymbol{\textnormal{\textbf{v}}}_{k}s_{k}}_{\text{Useful signal}}+\underbrace{\sum_{i\in\text{\ensuremath{\mathcal{K}}}_{\textnormal{dl}}\backslash k}\boldsymbol{\textnormal{\textbf{H}}}_{k,c_{i}}^{\textnormal{dl}}\boldsymbol{\textnormal{\textbf{v}}}_{i}s_{i}}_{\text{BS to UE interference}}+\underbrace{\sum_{j\in\text{\ensuremath{\mathcal{K}}}_{\textnormal{ul}}}\boldsymbol{\textnormal{\textbf{G}}}_{k,j}\boldsymbol{\textnormal{\textbf{w}}}_{j}s_{j}}_{\text{UE to UE interference}}+\boldsymbol{\textnormal{\textbf{n}}}_{k}^{\textnormal{dl}},
\end{equation}
}where $\boldsymbol{\textnormal{\textbf{H}}}_{c_{k},k}^{\textnormal{ul}}\in\text{\ensuremath{\mathcal{C}}}^{N\times M}$
denotes the 3GPP 3D-UMA spatial channel {[}4{]} from the $k^{th}$
UE to its $c_{k}^{th}$ BS serving BS, $\boldsymbol{\textnormal{\textbf{v}}}_{i}\in\text{\ensuremath{\mathcal{C}}}^{N\times1}$
, $\boldsymbol{\textnormal{\textbf{w}}}_{k}\in\text{\ensuremath{\mathcal{C}}}^{M\times1}$
and $s_{k}$ are the zero-forcing pre-coding vector at the $c_{i}^{th}$
BS, pre-coding vector of the the $k^{th}$ UE, and the transmitted
data symbol of the $k^{th}$ UE, respectively, while $\boldsymbol{\textnormal{\textbf{n}}}_{c_{k}}^{\textnormal{ul}}$
implies the additive white Gaussian noise.  Similarly, the UL signal
at the $c_{k}^{th}$ cell, $c_{k}\text{\ensuremath{\in\text{\ensuremath{\mathfrak{B}}}_{\textnormal{ul}}}}$
from $k\text{\ensuremath{\in\text{\ensuremath{\mathcal{K}}}_{\textnormal{ul}}}},$
is expressed by

{\small{}
\begin{equation}
\boldsymbol{\textnormal{y}}_{c_{k},k}^{\textnormal{ul}}=\underbrace{\boldsymbol{\textnormal{\textbf{H}}}_{c_{k},k}^{\textnormal{ul}}\boldsymbol{\textnormal{\textbf{w}}}_{k}s_{k}}_{\text{Useful signal}}+\underbrace{\sum_{j\in\text{\ensuremath{\mathcal{K}}}_{\textnormal{ul}}\backslash k}\boldsymbol{\textnormal{\textbf{H}}}_{c_{k},j}^{\textnormal{ul}}\boldsymbol{\textnormal{\textbf{w}}}_{j}s_{j}}_{\text{UE to BS interference}}+\underbrace{\sum_{i\in\text{\ensuremath{\mathcal{K}}}_{\textnormal{dl}}}\boldsymbol{\textnormal{\textbf{Q}}}_{c_{k},c_{i}}\boldsymbol{\textnormal{\textbf{v}}}_{i}s_{i}}_{\text{BS to BS interference}}+\boldsymbol{\textnormal{\textbf{n}}}_{c_{k}}^{\textnormal{ul}},
\end{equation}
}where $\boldsymbol{\textnormal{\textbf{Q}}}_{c_{k},c_{i}}\in\text{\ensuremath{\mathcal{C}}}^{N\times N}$
is the cross-link BS-BS channel between the serving BSs of the $k^{th}$
and $i^{th}$ UEs, $k\text{\ensuremath{\in\text{\ensuremath{\mathcal{K}}}_{\textnormal{ul}}}}$
and $i\in\text{\ensuremath{\mathcal{K}}}_{\textnormal{dl}}$. Then,
the post-receiver signal-to-interference ratio (SIR) in the DL $\gamma_{k}^{\textnormal{dl}}$
and UL $\gamma_{c_{k}}^{\textnormal{ul}}$ directions are given by,

{\small{}
\begin{equation}
\gamma_{k}^{\textnormal{dl}}=\frac{\left\Vert \left(\boldsymbol{\textnormal{\textbf{u}}}_{k}^{\textnormal{dl}}\right)^{\textnormal{H}}\boldsymbol{\textnormal{\textbf{H}}}_{k,c_{k}}^{\textnormal{dl}}\boldsymbol{\textnormal{\textbf{v}}}_{k}\right\Vert ^{2}}{\underset{i\in\text{\ensuremath{\mathcal{K}}}_{\textnormal{dl}}\backslash k}{\sum}\left\Vert \left(\boldsymbol{\textnormal{\textbf{u}}}_{k}^{\textnormal{dl}}\right)^{\textnormal{H}}\boldsymbol{\textnormal{\textbf{H}}}_{k,c_{i}}^{\textnormal{dl}}\boldsymbol{\textnormal{\textbf{v}}}_{i}\right\Vert ^{2}+\underset{j\in\text{\ensuremath{\mathcal{K}}}_{\textnormal{ul}}}{\sum}\left\Vert \left(\boldsymbol{\textnormal{\textbf{u}}}_{k}^{\textnormal{dl}}\right)^{\textnormal{H}}\boldsymbol{\textnormal{\textbf{G}}}_{k,j}\boldsymbol{\textnormal{\textbf{w}}}_{j}\right\Vert ^{2}},
\end{equation}
}{\small \par}

{\small{}
\begin{equation}
\gamma_{c_{k}}^{\textnormal{ul}}=\frac{\left\Vert \left(\boldsymbol{\textnormal{\textbf{u}}}_{k}^{\textnormal{ul}}\right)^{\textnormal{H}}\boldsymbol{\textnormal{\textbf{H}}}_{c_{k},k}^{\textnormal{ul}}\boldsymbol{\textnormal{\textbf{w}}}_{k}\right\Vert ^{2}}{\underset{j\in\text{\ensuremath{\mathcal{K}}}_{\textnormal{ul}}\backslash k}{\sum}\left\Vert \left(\boldsymbol{\textnormal{\textbf{u}}}_{k}^{\textnormal{ul}}\right)^{\textnormal{H}}\boldsymbol{\textnormal{\textbf{H}}}_{c_{k},j}^{\textnormal{ul}}\boldsymbol{\textnormal{\textbf{w}}}_{j}\right\Vert ^{2}+\underset{i\in\text{\ensuremath{\mathcal{K}}}_{\textnormal{dl}}}{\sum}\left\Vert \left(\boldsymbol{\textnormal{\textbf{u}}}_{k}^{\textnormal{ul}}\right)^{\textnormal{H}}\boldsymbol{\textnormal{\textbf{Q}}}_{c_{k},c_{i}}\boldsymbol{\textnormal{\textbf{v}}}_{i}\right\Vert ^{2}},
\end{equation}
}where $\left\Vert \bullet\right\Vert ^{2}$ is the second-norm, $\boldsymbol{\textnormal{\textbf{u}}}_{k}^{\kappa}\in\text{\ensuremath{\mathcal{C}}}^{N/M\times1}$,
$\text{\ensuremath{\mathcal{X}}}^{\kappa},\kappa\text{\ensuremath{\in}}\{\textnormal{ul},\textnormal{dl}\}$,
is the linear minimum mean square error interference rejection combining
(LMMSE-IRC) receiver vector {[}5{]}, with $\left(\bullet\right)^{\textnormal{H}}$
as the Hermitian operation. 
\begin{figure}
\begin{centering}
\includegraphics[scale=0.35]{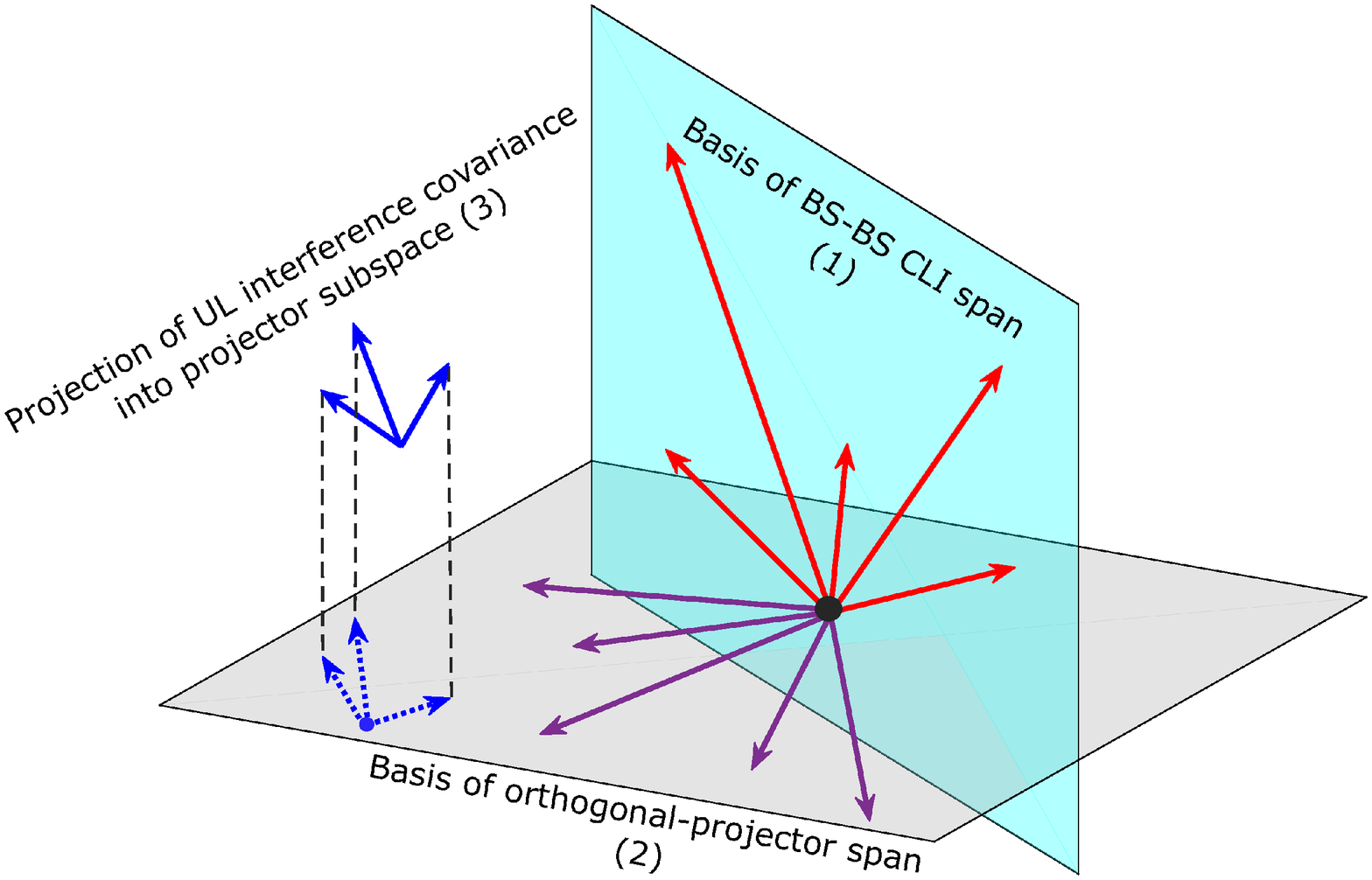}
\par\end{centering}
\centering{}{\small{}Fig. 2. BS-BS CSA: CLI projection onto projector
sub-space.}{\small \par}
\end{figure}

\section{Proposed BS-BS CLI suppression algorithm }

The proposed CSA offers an efficient BS-BS CLI cancellation with a
limited and 3GPP-compliant overhead space. First, based on {[}4{]},
the UE-UE CLI is reliably pre-avoided. Then, during the BS-BS CLI
slots, victim UL BSs identify the basis of the principal BS-BS CLI
interfering sub-space using a \textit{DL precoder map} signaling over
the \textit{Xn-interface}. Then, UL BSs estimate the corresponding
orthonormal projector sub-space. Finally, for every impacted UL transmission,
UL BSs spatially project the estimated IRC interference covariance
onto the projector sub-space, prior to decoding, as shown in Fig.
2.

\subsection{Link-direction adaptation }

During each RFC update instance, each BS independently selects an
RFC from the RFC code-book which best satisfies its individual link-direction
selection criterion, with a respective DL-to-UL symbol ratio, i.e.,
$d_{c}:u_{c}$. We adopt the DL and UL buffered traffic size as the
main criterion to select an RFC. The buffered traffic ratio $\mu_{c}\left(t\right)$
is defined as

\begin{equation}
\mu_{c}\left(t\right)=\frac{Z_{c}^{\textnormal{dl}}\left(t\right)}{Z_{c}^{\textnormal{dl}}\left(t\right)+Z_{c}^{\textnormal{ul}}\left(t\right)},
\end{equation}
where $Z_{c}^{\textnormal{dl}}\left(t\right)$ and $Z_{c}^{\textnormal{ul}}\left(t\right)$
are the total buffered DL and UL traffic of the $c_{k}^{th}$ BS at
the RFC update time $t$. For example, at the $c^{th}$ BS with $\mu_{c}\left(t\right)=0.3$,
the buffered UL traffic volume is $\textnormal{2.3x}$ the buffered
DL traffic, thus, BS consequently selects a slot format of DL:UL symbol
ratio as $\sim1:2.3$. The placement of the DL and UL symbols during
a slot duration is set evenly to allow for multiple scattered DL and
UL transmission opportunities. Accordingly, the achievable capacity
$T$ of each cluster is given by

\begin{equation}
T=\stackrel[c=1]{C}{\sum}\min\left(\textnormal{\ensuremath{u_{c}},\ensuremath{u_{c}^{\textnormal{opt.}}}}\right)\digamma_{c}^{u}+\min\left(\textnormal{\ensuremath{d_{c}},\ensuremath{d_{c}^{\textnormal{opt.}}}}\right)\digamma_{c}^{d},
\end{equation}
where $\digamma_{c}^{u}$ and $\digamma_{c}^{d}$ represent the rate
utility functions of the UL and DL directions, respectively. $u_{c}^{\textnormal{opt.}}$
and $d_{c}^{\textnormal{opt.}}$ are the optimal numbers of the UL
and DL slots that should be adopted during the current RFC to perfectly
match the current traffic variations. Thus, UL $\chi^{\textnormal{ul}}$
and DL $\chi^{\textnormal{dl}}$ symbol mismatch are inflicted due
to the insufficient RFC quantization as 

\begin{equation}
\chi^{\textnormal{ul}}=\left|\ensuremath{u_{c}}-\ensuremath{u_{c}^{\textnormal{opt.}}}\right|.
\end{equation}

\begin{equation}
\chi^{\textnormal{dl}}=\left|\ensuremath{d_{c}}-\ensuremath{d_{c}^{\textnormal{opt.}}}\right|.
\end{equation}

To maximize capacity $T$, $u_{c}=u_{c}^{\textnormal{opt.}}$ and
$d_{c}=d_{c}^{\textnormal{opt.}}$ should be always satisfied. Although,
$u_{c}^{\textnormal{opt.}}$ and $d_{c}^{\textnormal{opt.}}$ may
introduce severe BS-BS CLI which severely degrades the UL capacity. 

\subsection{Proposed BS-BS CSA}

During the inter-BS CLI slots within an RFC, the DL-aggressor BSs
signal adjacent victim UL BSs with a \textit{DL precoder map} over
the \textit{Xn-interface}. Such on-demand signaling denotes a vector
of the DL sub-band pre-coding matrix indices (PMIs), which will be
used during the next slot by the scheduled DL users. For instance,
with 10 MHz bandwidth, i.e., 50 PRBs, $4$ antenna port setup, i.e.,
4-bit PMI, 3 BS-BS CLI slots, 8-PRB sub-bands, the size of the \textit{DL
precoder map} $\text{\ensuremath{\mathbb{O}}}$ can be calculated
as: 

\begin{equation}
\text{\ensuremath{\mathbb{O}}}=3\times\left(\frac{50}{8}\times\left(\log_{2}\left(\frac{50}{8}\right)+4\right)\right)\simeq124\,\textnormal{ bits per 10 ms}.
\end{equation}

Accordingly, the victim UL BSs seek to identify the strongest $N-1$
sub-band BS-BS interferers as

\begin{equation}
\text{\ensuremath{\mathit{\Lambda}}}_{b_{\textnormal{ul}},b_{\textnormal{dl}}}^{l}=\left\Vert \boldsymbol{\textnormal{\textbf{Q}}}_{b_{\textnormal{ul}},b_{\textnormal{dl}}}^{l}\boldsymbol{\textnormal{\textbf{v}}}_{b_{\textnormal{dl}}}^{l}\right\Vert ^{2},\,\,\,b_{\textnormal{dl}}\in\text{\ensuremath{\mathfrak{B}}}_{\textnormal{dl}},b_{\textnormal{ul}}\in\text{\ensuremath{\mathfrak{B}}}_{\textnormal{ul}},l\in L
\end{equation}

\begin{equation}
\left(\text{\ensuremath{\mathfrak{J}} }_{1}^{b_{\textnormal{ul}},l},\ldots,\text{\ensuremath{\mathfrak{J}} }_{N-1}^{b_{\textnormal{ul}},l}\right)=\left\{ \boldsymbol{\textnormal{\textbf{Q}}}_{b_{\textnormal{ul}},b_{\textnormal{dl}}}^{l}\boldsymbol{\textnormal{\textbf{v}}}_{b_{\textnormal{dl}}}^{l}\rightarrow\underset{b_{\textnormal{dl}},l}{\arg\max}\left(\text{\ensuremath{\mathit{\Lambda}}}_{b_{\textnormal{ul}},b_{\textnormal{dl}}}^{l}\right)\right\} ,
\end{equation}
where $\boldsymbol{\textnormal{\textbf{Q}}}_{b_{\textnormal{ul}},b_{\textnormal{dl}}}^{l}$
is the BS-BS channel between the $b_{\textnormal{ul}}^{th}$ and $b_{\textnormal{dl}}^{th}$
BSs over the $l^{th}$ sub-band, with $L$ as the number of DL aggressor
sub-bands. $\boldsymbol{\textnormal{\textbf{v}}}_{b_{\textnormal{dl}}}^{l}$
implies the DL precoder of the scheduled user over the $l^{th}$ sub-band
at the $b_{\textnormal{dl}}^{th}$ BS, and $\text{\ensuremath{\mathfrak{J}} }_{i}^{b_{\textnormal{ul}},l}\in\text{\ensuremath{\mathcal{C}}}^{N\times1}$,
with $i=1,2,\ldots N-1,$ are the identified strongest BS-BS interfering
vectors at the $b_{\textnormal{ul}}^{th}$ BS. Since the strongest
BS-BS interferers, i.e., $\boldsymbol{\textnormal{\textbf{Q}}}_{b_{\textnormal{ul}},b_{\textnormal{dl}}}^{l}\boldsymbol{\textnormal{\textbf{v}}}_{b_{\textnormal{dl}}}^{l},$
are linearly independent due to the independent inter-cell user scheduling,
we can utilize the Gram Schmidt orthogonalization {[}14{]} for victim
UL BSs to estimate the basis vectors $\beta_{i}^{b_{\textnormal{ul}},l}\in\text{\ensuremath{\mathcal{C}}}^{N\times1}$
of a spatial sub-space that spans all $N-1$ BS-BS interferers, as 

\begin{equation}
\beta_{i}^{b_{\textnormal{ul}},l}=\left\{ \begin{array}{c}
\text{\ensuremath{\mathfrak{J}} }_{1}^{b_{\textnormal{ul}},l},\,\,\,\,\,i=1\\
\text{\ensuremath{\mathfrak{J}} }_{i}^{b_{\textnormal{ul}},l}-\stackrel[\tau=1]{i-1}{\sum}\textnormal{pro}\textnormal{j}_{\beta_{\tau}}\left(\text{\ensuremath{\mathfrak{J}} }_{i}^{b_{\textnormal{ul}},l}\right),\,\,\,\,\,2\geq i\leq N-1,
\end{array}\right.
\end{equation}

\begin{equation}
\textnormal{pro}\textnormal{j}_{\beta_{\tau}}\left(\text{\ensuremath{\mathfrak{J}} }_{i}^{b_{\textnormal{ul}},l}\right)=\left(\frac{\mathfrak{J}_{i}^{b_{\textnormal{ul}},l}\,.\,\beta_{\tau}}{\left\Vert \beta_{\tau}\right\Vert ^{2}}\right)\beta_{\tau},
\end{equation}
where $\textnormal{pro}\textnormal{j}_{\text{\ensuremath{\mathsf{X}}}}\left(\text{\ensuremath{\mathsf{Y}}}\right)$
implies the spatial line-projection of vector $\text{\ensuremath{\mathsf{Y}}}$
on vector $\text{\ensuremath{\mathsf{X}}},$ while $\left(\mathsf{X}\,.\,\mathsf{Y}\right)$
is the dot product. Then, the BS-BS CLI basis matrix $\text{\ensuremath{\mathcal{A}}}\in\text{\ensuremath{\mathcal{C}}}^{N\times N-1}$
is constructed as

\begin{equation}
\text{\ensuremath{\mathcal{A}}}=\left[\beta_{1}^{b_{\textnormal{ul}},l},\,\beta_{2}^{b_{\textnormal{ul}},l},\,\ldots\,,\,\beta_{N-1}^{b_{\textnormal{ul}},l}\right].
\end{equation}

The UL BSs accordingly estimate an orthonormal projector subspace
$\text{\ensuremath{\mathcal{A}}}^{\text{\textSFvii}}\in\text{\ensuremath{\mathcal{C}}}^{N\times N}$
by the orthogonal projection, as

\begin{equation}
\text{\ensuremath{\mathcal{A}}}^{\text{\textSFvii}}=\text{\ensuremath{\mathcal{A}}}\left(\text{\ensuremath{\mathcal{A}}}^{\textnormal{T}}\text{\ensuremath{\mathcal{A}}}\right)^{-1}\text{\ensuremath{\mathcal{A}}}^{\textnormal{T}},
\end{equation}
where $\left(\bullet\right)^{\textnormal{-1}}$ and $\left(\bullet\right)^{\textnormal{T}}$
are the inverse and transpose operations. Finally, for each UL transmission
during the current BS-BS CLI slot, UL BSs calculate the average UL
interference covariance matrix $\textnormal{\textbf{R}}_{k}^{\textnormal{ul}}\in\text{\ensuremath{\mathcal{C}}}^{N\times N},$
in order to construct the LMMSE-IRC receiver matrix for decoding,
expressed as 

\begin{equation}
\Xi_{k}^{\textnormal{ul}}=\underbrace{\sum_{j\in\text{\ensuremath{\mathcal{K}}}_{\textnormal{ul}}\backslash k}\boldsymbol{\textnormal{\textbf{H}}}_{c_{k},j}^{\textnormal{ul}}\boldsymbol{\textnormal{\textbf{w}}}_{j}}_{\text{Same-link}}+\underbrace{\sum_{i\in\text{\ensuremath{\mathcal{K}}}_{\textnormal{dl}}}\boldsymbol{\textnormal{\textbf{Q}}}_{c_{k},c_{i}}\boldsymbol{\textnormal{\textbf{v}}}_{i}}_{\text{Cross-link}}.
\end{equation}

\begin{equation}
\textnormal{\textbf{R}}_{k}^{\textnormal{ul}}=\Xi_{k}^{\textnormal{ul}}\times\left(\Xi_{k}^{\textnormal{ul}}\right)^{\textnormal{H}}.
\end{equation}

Such interference estimate is highly sparse in the spatial domain
due to the BS-BS CLI summation, leading to a degraded linear-IRC decoding
performance. Thus, prior to decoding, the UL BSs spatially project
the interference column vectors of $\textnormal{\textbf{R}}_{k}^{\textnormal{ul}}$,
i.e., $\mathit{\textnormal{\textbf{r}}}_{\rho}^{\textnormal{ul}}$,
onto the projector sub-space basis as

\begin{equation}
\breve{\boldsymbol{\textnormal{\textbf{r}}}}_{\rho}^{\textnormal{ul}}=\textnormal{pro}\textnormal{j}_{_{\boldsymbol{\textnormal{\textbf{a}}}_{\rho}^{\text{\textSFvii}}}}\left(\mathit{\textnormal{\textbf{r}}}_{\rho}^{\textnormal{ul}}\right)=\frac{\mathit{\textnormal{\textbf{r}}}_{\rho}^{\textnormal{ul}}\,.\,\boldsymbol{\textnormal{\textbf{a}}}_{\rho}^{\text{\textSFvii}}}{\left\Vert \boldsymbol{\boldsymbol{\textnormal{\textbf{a}}}}_{\rho}^{\text{\textSFvii}}\right\Vert ^{2}}\times\boldsymbol{\textnormal{\textbf{a}}}_{\rho}^{\text{\textSFvii}},\,\,\,\,\,\forall\rho=1,2,\ldots,N.
\end{equation}
with $\boldsymbol{\textnormal{\textbf{a}}}_{\rho}^{\text{\textSFvii}}$
and $\breve{\boldsymbol{\textnormal{\textbf{r}}}}_{\rho}^{\textnormal{ul}}$
are the column vectors of the projector sub-space $\text{\ensuremath{\mathcal{A}}}^{\text{\textSFvii}}$
and the updated interference covariance matrix $\breve{\boldsymbol{\textnormal{\textbf{R}}}}_{k}^{\textnormal{ul}}$.
Hence, the spatial span of $\breve{\boldsymbol{\textnormal{\textbf{R}}}}_{k}^{\textnormal{ul}}$
is regularized by suppressing the sparse $N-1$ BS-BS CLI strongest
aggressors, i.e., $\sim$ removing the second summation of eq. (16).
Finally, the UL LMMSE-IRC receiver matrix is then designed as 

\begin{equation}
\boldsymbol{\textnormal{\textbf{u}}}_{k}^{\textnormal{ul}}=\left(\boldsymbol{\textnormal{\textbf{H}}}_{c_{k},k}^{\textnormal{ul}}\boldsymbol{\textnormal{\textbf{w}}}_{k}\left(\boldsymbol{\textnormal{\textbf{H}}}_{c_{k},k}^{\textnormal{ul}}\boldsymbol{\textnormal{\textbf{w}}}_{k}\right)^{\textnormal{H}}+\breve{\boldsymbol{\textnormal{\textbf{R}}}}_{k}^{\textnormal{ul}}\right)^{^{-1}}\,\boldsymbol{\textnormal{\textbf{H}}}_{c_{k},k}^{\textnormal{ul}}\boldsymbol{\textnormal{\textbf{w}}}_{k}.
\end{equation}

Therefore, the UL decoder becomes highly directive towards the span
of the direct effective channel, and outside the subspace spanned
by the principal BS-BS CLI basis, leading to a significant improvement
of the URLLC UL performance. 

\section{Simulation Results }

We adopt extensive system-level simulations to evaluate the performance
of the proposed BS-BS CSA, where the major 3GPP 5G-NR assumptions
for URLLC {[}4{]} are followed, and as listed in Table I. A $8\times2$
antenna setup along with 10 MHz bandwidth of 30 KHz SCS are configured,
while the DL transmission power is set to 40 dBm. The offered DL traffic
is set to 2x times the UL traffic. During each TTI, BSs dynamically
schedule active UEs using the proportional fair (PF) criterion. The
achievable SC SINRs are combined using the exponential SNR mapping
{[}15{]} in order to estimate an effective SINR level. Accordingly,
fully dynamic modulation and coding selection (MCS) and Chase combining
HARQ re-transmissions are utilized. Pending HARQ re-transmissions
are always prioritized over new transmissions during the first available
DL/UL slot transmission opportunity. We assess the performance of
the proposed solution against the state-of-the-art dynamic-TDD studies
as follows: 
\begin{table}
\caption{{\small{}Default simulation parameters.}}
\centering{}%
\begin{tabular}{c|c}
\hline 
Parameter & Value\tabularnewline
\hline 
Environment & 3GPP-UMA, one cluster, 21 cells\tabularnewline
\hline 
UL/DL channel bandwidth & 10 MHz, SCS = 30 KHz, TDD\tabularnewline
\hline 
TDD mode & Synchronized\tabularnewline
\hline 
Antenna setup & $N=4$, $M=4$\tabularnewline
\hline 
UL power control & $\alpha=1,\:P0=-103$ dBm\tabularnewline
\hline 
Link adaptation & Adaptive modulation and coding \tabularnewline
\hline 
HARQ configuration & Asynchronous, Chase Combining\tabularnewline
\hline 
Processing times & $\begin{array}{c}
\textnormal{PDSCH : 4.5-OFDM symbols}\\
\textnormal{PUSCH : 5.5-OFDM symbols}
\end{array}$\tabularnewline
\hline 
TTI configuration & 4-OFDM symbols\tabularnewline
\hline 
Traffic model & $\begin{array}{c}
\textnormal{\textnormal{FTP3}}\\
\textnormal{\textnormal{\ensuremath{\mathit{f}^{dl}}} = \textnormal{\ensuremath{\mathit{f}^{ul}}} = 400 bits}
\end{array}$\tabularnewline
\hline 
Offered traffic ratio & DL:UL = 2 : 1\tabularnewline
\hline 
DL/UL scheduling & Proportional fair\tabularnewline
\hline 
DL/UL receiver & LMMSE-IRC\tabularnewline
\hline 
Pattern update periodicity & Slot duration\tabularnewline
\hline 
Transport layer setup & UDP, MTU = 1500 Bytes\tabularnewline
\hline 
User scheduler & Proportional fair\tabularnewline
\hline 
\end{tabular}
\end{table}
 
\begin{table*}
\caption{{\small{}Comparison of the URLLC outage latency, with offered load
per BS, and DL:UL = $2:1$.}}
\centering{}%
\begin{tabular}{c|c|c|c|c|c|c|c|c}
\hline 
\multirow{2}{*}{$\begin{array}{c}
\textnormal{Offered}\\
\textnormal{load}
\end{array}$} & \multicolumn{2}{c|}{CF-TDD} & \multicolumn{2}{c|}{NC-TDD} & \multicolumn{2}{c|}{CRFC-TDD} & \multicolumn{2}{c}{Proposed BS-BS CSA}\tabularnewline
\cline{2-9} 
 & DL  & UL & DL  & UL & DL  & UL & DL  & UL\tabularnewline
\hline 
4 Mbps & $\begin{array}{c}
7.15\\
\hilight{0.0\%}
\end{array}$ & $\begin{array}{c}
14.76\\
\hilight{0.0\%}
\end{array}$ & $\begin{array}{c}
8.47\\
\hilight{+16.9\%}
\end{array}$ & $\begin{array}{c}
105.34\\
\hilight{+150.8\%}
\end{array}$ & $\begin{array}{c}
7.75\\
\hilight{+8.0\%}
\end{array}$ & $\begin{array}{c}
24.12\\
\hilight{+48.1\%}
\end{array}$ & $\begin{array}{c}
7.36\\
\hilight{+2.89\%}
\end{array}$ & $\begin{array}{c}
17.4\\
\hilight{+16.4\%}
\end{array}$\tabularnewline
\hline 
5 Mbps & $\begin{array}{c}
8.04\\
\hilight{0.0\%}
\end{array}$ & $\begin{array}{c}
15.17\\
\hilight{0.0\%}
\end{array}$ & $\begin{array}{c}
1663\\
\hilight{+198\%}
\end{array}$ & $\begin{array}{c}
6063\\
\hilight{+199\%}
\end{array}$ & $\begin{array}{c}
14.24\\
\hilight{+55.6\%}
\end{array}$ & $\begin{array}{c}
201.6\\
\hilight{+172\%}
\end{array}$ & $\begin{array}{c}
8.43\\
\hilight{+4.7\%}
\end{array}$ & $\begin{array}{c}
18.0\\
\hilight{+17\%}
\end{array}$\tabularnewline
\hline 
6 Mbps & $\begin{array}{c}
11.04\\
\hilight{0.0\%}
\end{array}$ & $\begin{array}{c}
16.29\\
\hilight{0.0\%}
\end{array}$ & $\begin{array}{c}
7394\\
\hilight{+199.4\%}
\end{array}$ & $\begin{array}{c}
18390\\
\hilight{+199.6\%}
\end{array}$ & $\begin{array}{c}
3150\\
\hilight{+198.6\%}
\end{array}$ & $\begin{array}{c}
12540\\
\hilight{+199.4\%}
\end{array}$ & $\begin{array}{c}
11.47\\
\hilight{+3.82\%}
\end{array}$ & $\begin{array}{c}
19.32\\
\hilight{+17\%}
\end{array}$\tabularnewline
\hline 
7 Mbps & $\begin{array}{c}
17.28\\
\hilight{0.0\%}
\end{array}$ & $\begin{array}{c}
18.23\\
\hilight{0.0\%}
\end{array}$ & $\begin{array}{c}
12480\\
\hilight{+199.4\%}
\end{array}$ & $\begin{array}{c}
25610\\
\hilight{+199.7\%}
\end{array}$ & $\begin{array}{c}
6575\\
\hilight{+198.9\%}
\end{array}$ & $\begin{array}{c}
19470\\
\hilight{+199.6\%}
\end{array}$ & $\begin{array}{c}
19.8\\
\hilight{+13.5\%}
\end{array}$ & $\begin{array}{c}
23.07\\
\hilight{+23.4\%}
\end{array}$\tabularnewline
\hline 
\end{tabular}
\end{table*}

\textbf{CLI-free TDD} (CF-TDD) {[}6{]}: a fully dynamic TDD setup,
where BSs independently select the RFCs that best meet their individual
traffic demands; however, with the assumption of a perfect UE-UE and
BS-BS CLI cancellation. We consider such optimal; although, theoretical
baseline, as the reference case.

\textbf{Non-coordinated TDD} (NC-TDD): a fully dynamic TDD is assumed;
however, neither inter-BS coordination nor UE-UE and BS-BS CLI cancellation
are supported. Herein, BSs achieve the maximum dynamic-TDD adaptation;
though, with potentially severe BS-BS and UE-UE CLI, respectively. 

\textbf{Coordinated-RFC TDD} (CRFC-TDD) {[}4{]}: a hybrid frame design
along with a cyclic-offset-based RFC code-book are constructed to
reliably pre-avoid the UE-UE CLI. That is, UEs with the worst radio
conditions, are preemptively scheduled during certain CLI-free slots,
i.e., static slots within all RFCs. Hence, CRFC-TDD boosts the cell-edge
capacity; though, performance is highly limited by the more critical
BS-BS CLI.

\begin{figure}
\begin{centering}
\includegraphics[scale=0.55]{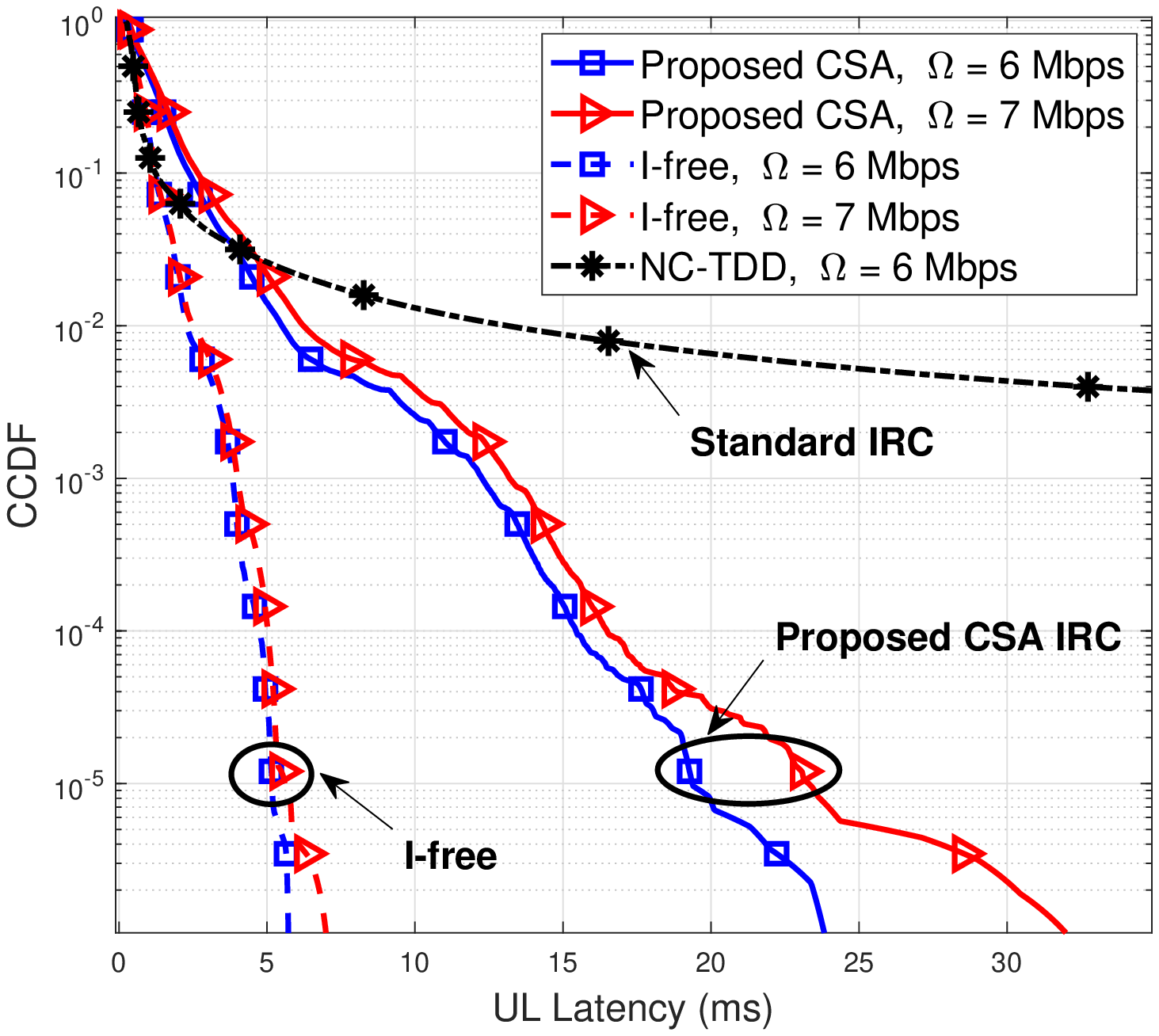}
\par\end{centering}
\centering{}{\small{}Fig. 3. BS-BS CSA: UL latency performance. }{\small \par}
\end{figure}
 
\begin{figure}
\begin{centering}
\includegraphics[scale=0.55]{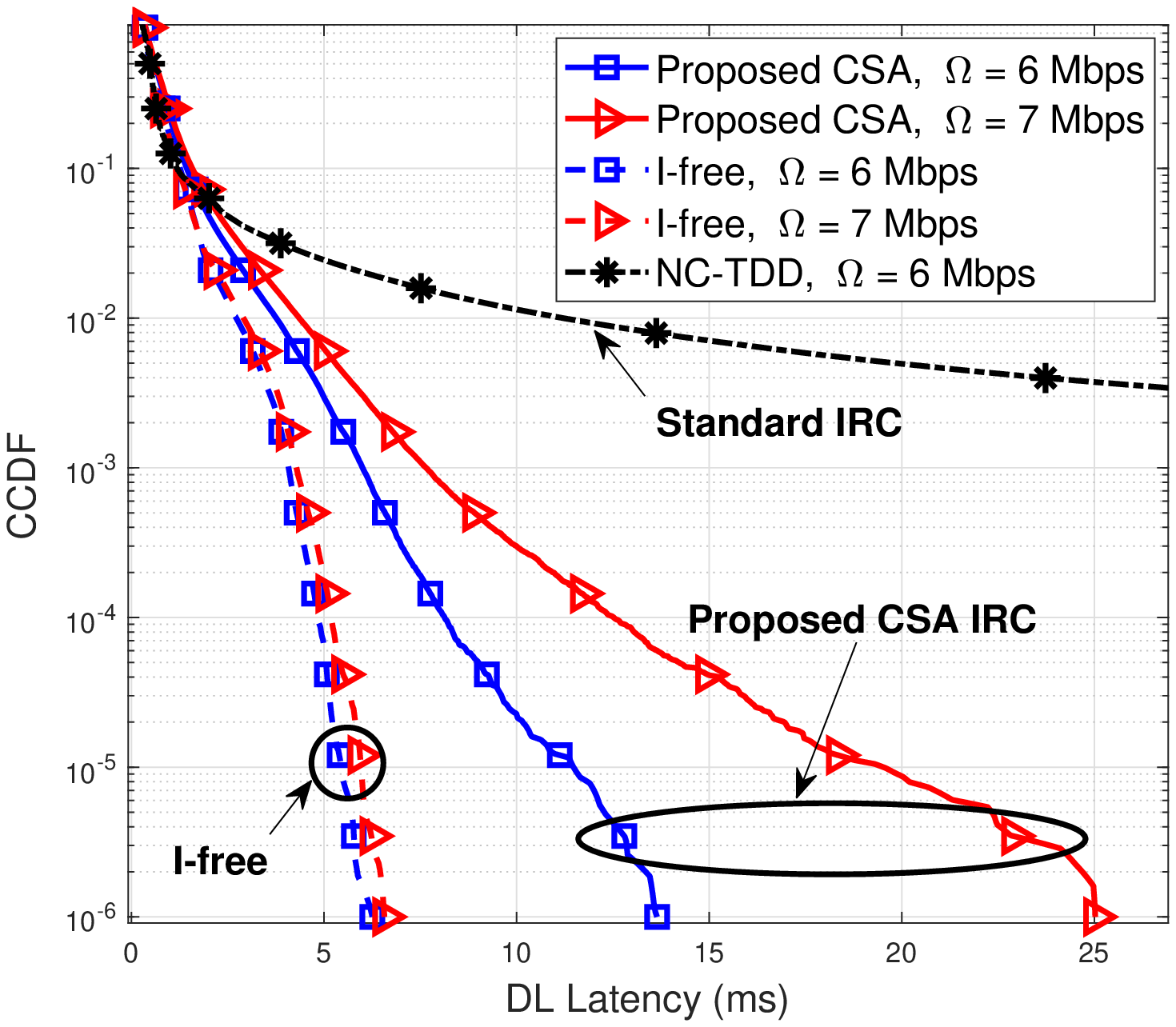}
\par\end{centering}
\centering{}{\small{}Fig. 4. BS-BS CSA: DL latency performance. }{\small \par}
\end{figure}
We first evaluate the performance of the proposed scheme in terms
of the URLLC outage latency. That is, the achievable URLLC radio latency
at $10^{-5}$ outage probability. It implies the one-way radio latency
from the moment a packet arrives at transmitter until it has been
successfully decoded at the receiver end, including the standard BS
and UE processing delays, dynamic user scheduling delay, and the HARQ
re-transmission buffering delay, respectively. Thus, Fig. 3 and 4
depict the complementary cumulative distribution function (CCDF) of
the UL and DL URLLC latency, respectively, under various offered load
levels for the proposed CSA, NC-TDD, and the hypothetical; though,
optimal, interference-free (I-free) case, where we assume a perfect
inter-cell interference cancellation, including the same-link and
cross-link interference. As clearly shown, the proposed CSA scheme
offers a decent URLLC outage latency due to the enhanced suppression
of the principal BS-BS CLI interferers. The degraded outage latency
under the high offered load region is attributed to the inflicted
queuing delay due to the dynamic user scheduling, and the increasing
same-link inter-cell interference. The NC-TDD with the standard IRC
receiver design clearly inflicts a significant degradation of the
achievable URLLC latency due to the severe BS-BS CLI. 

\begin{figure}
\begin{centering}
\includegraphics[scale=0.62]{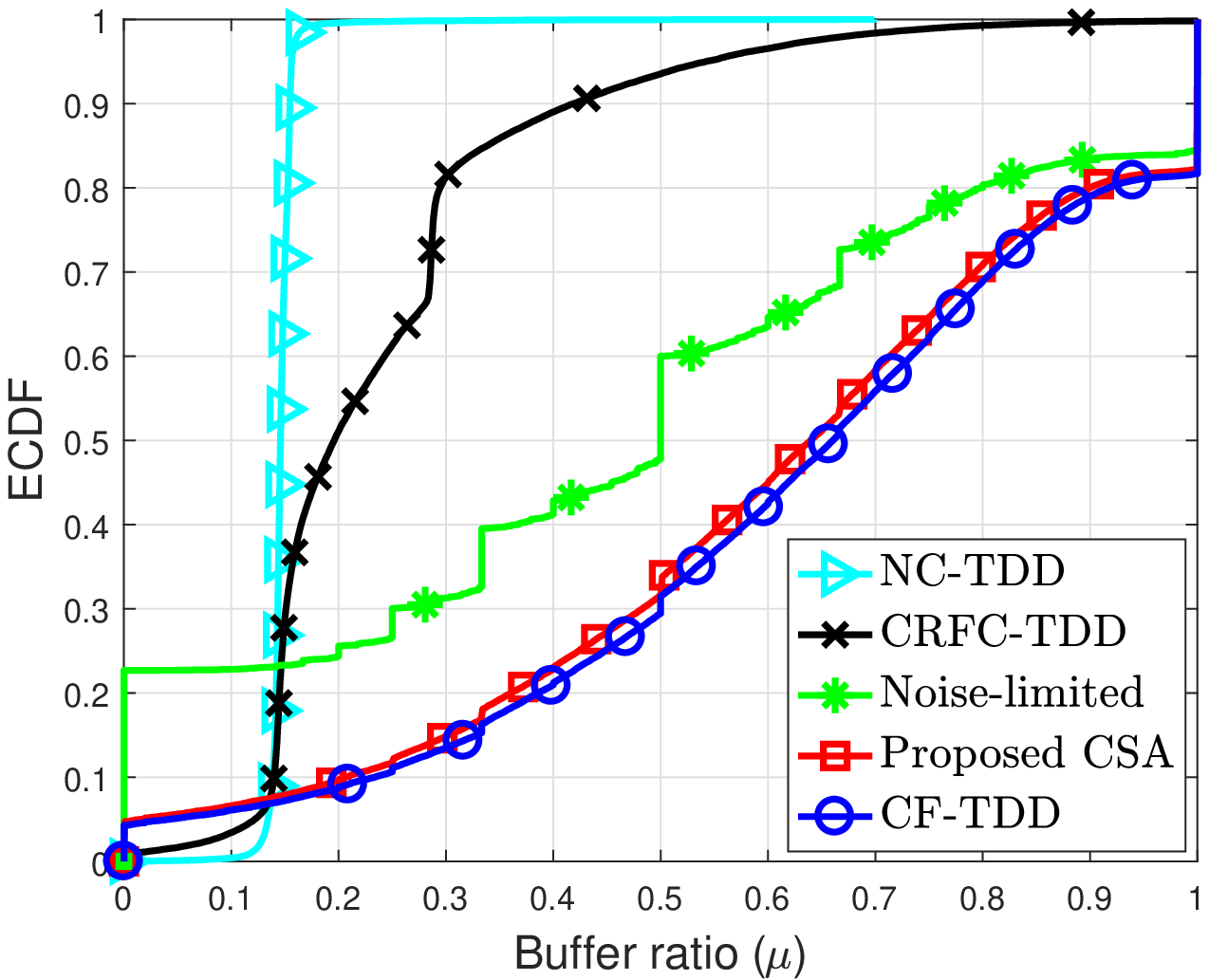}
\par\end{centering}
\centering{}{\small{}Fig. 5. BS-BS CSA: traffic buffering performance. }{\small \par}
\end{figure}
Table II holds a comparison of the URLLC radio latency in ms, for
all schemes under evaluation at different offered traffic loads per
BS. To reflect the URLLC reliability targets, the URLLC outage latency
at the $10^{-5}$ outage probability level is evaluated. The CF-TDD
clearly provides the best URLLC outage latency performance due to
the absolute absence of the UE-UE and BS-BS CLI. The NC-TDD and CRFC-TDD
schemes fail to offer a decent URLLC DL and UL outage latency, mainly
due to the severe and unhandled BS-BS CLI. Under high offered loads,
their respective outage latency increases dramatically due to the
inflicted UL re-transmissions.

\begin{figure}
\begin{centering}
\includegraphics[scale=0.62]{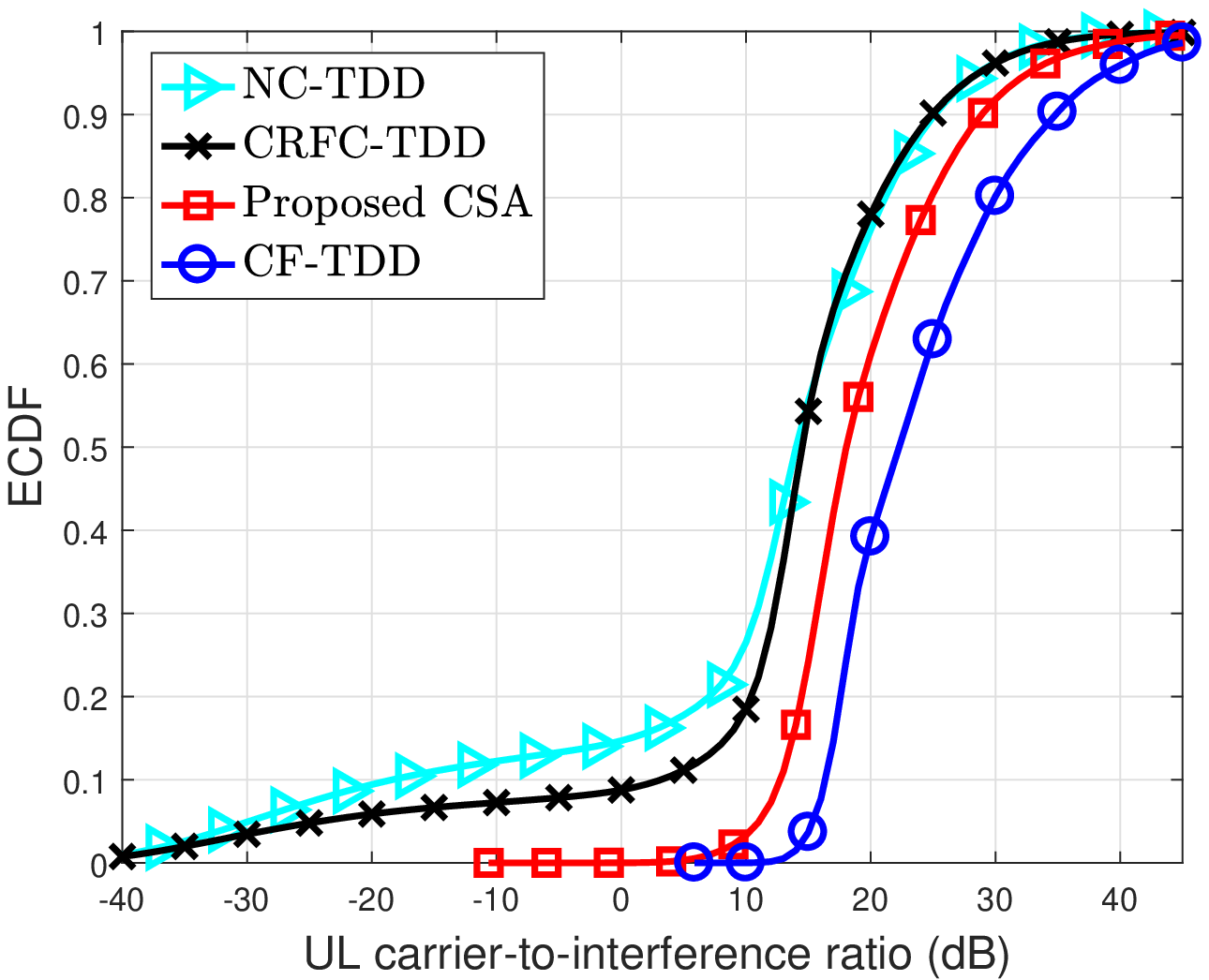}
\par\end{centering}
\centering{}{\small{}Fig. 6. BS-BS CSA: UL interference performance.}{\small \par}
\end{figure}
The proposed BS-BS CSA offers a significant improvement of the URLLC
DL and UL outage latency, clearly approaching the optimal CF-TDD under
all offered loads; however, with a significantly reduced control overhead
size. Due to the sufficient BS-BS CLI suppression, the proposed solution
guarantees faster UL transmissions without several HARQ re-transmissions,
leaving more time and resources for DL traffic.

These conclusions are confirmed by examining the empirical CDF (ECDF)
of the buffered traffic ratio $\mu$ as in eq. (5), and shown by Fig.
5. The lower $\mu$, the higher the buffered UL traffic in the scheduling
queues. Herein, we introduce a hypothetical case, where the system
is only noise-limited, i.e., inter-cell same-link and cross-link interference
is assumed to be perfectly suppressed (I-free case as depicted by
Fig. 3 and 4). This case provides a fairer buffer ratio, i.e., $\mu=0.5$
at the $\textnormal{50}$ percentile since all DL and UL payloads
get successfully decoded from the first time. The NC-TDD and CRFC-TDD
offer an extremely low $\mu$, i.e., $\mu=0.15$ and $0.2$ at the
$\textnormal{50}$ percentile. That is, the buffered UL traffic is
$5.6$x and $4$x times the buffered DL traffic, respectively, despite
that the offered DL traffic is twice the offered UL traffic. This
is due to the UL traffic excessive buffering, due to the consistent
consumption of the maximum UL HARQ attempts before failure, and caused
by the severe BS-BS CLI. This denotes the link direction adaptation
of the dynamic TDD becomes dictated by the HARQ performance, rather
than by the new packet arrivals. However, the proposed BS-BS CSA and
optimal CF-TDD offer a smooth buffering performance, i.e., $\mu=0.66,$
which implies that buffered UL traffic is $0.525$x times the buffered
DL traffic, that perfectly aligns with the configured offered traffic
ratio. 

Fig. 6 presents the ECDF of the UL carrier-to-interference ratio (CIR)
in dB. For a proper presentation, the artificial noise-limited case
is excluded. The NC-TDD obviously exhibits the worst CIR performance.
The CRFC-TDD only outperforms the NC-TDD over the lower percentiles
(cell edge UEs), i.e., $+22$ dB increase at the $\textnormal{10}$
percentile, due to the reliable UE-UE CLI pre-avoidance. Proposed
solution offers $+31$ dB and $+9$ dB CIR improvements at the $\textnormal{10}$
percentile, compared to the NC-TDD, and CRFC-TDD, respectively. Unlike
the CRFC-TDD, the CIR gain of the proposed solution does not vanish
over the higher percentiles, due to the sufficient BS-BS CLI suppression.
Proposed scheme approaches the optimal CF-TDD with an average loss
of $-4$ dB. 

Similar conclusions are also drawn from Fig. 7, where the ECDF of
the UL throughput per packet is depicted. At the $\textnormal{10}$
percentile, the proposed BS-BS CSA offers $\sim+156\%$ increase in
the achievable URLLC packet throughput, compared to the NC-TDD scheme.
This is mainly attributed to the achievable CIR gain of the proposed
CSA solution. 
\begin{figure}
\begin{centering}
\includegraphics[scale=0.62]{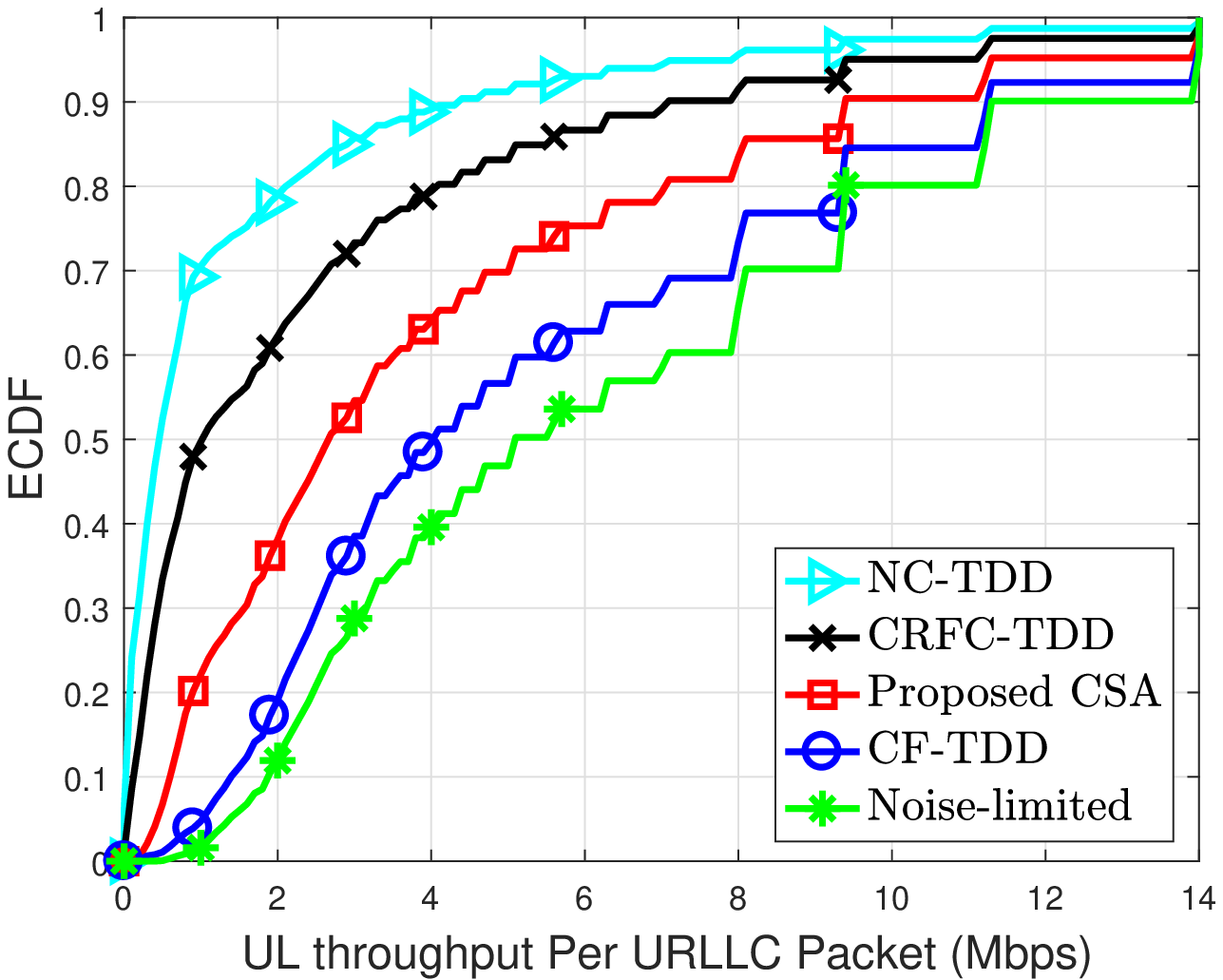}
\par\end{centering}
\centering{}{\small{}Fig. 7. BS-BS CSA: UL packet throughput performance.}{\small \par}
\end{figure}

\section{Concluding Remarks }

A high-performance and computation-efficient cross-link interference
(CLI) suppression algorithm has been proposed in this work, for 5G
dynamic-TDD macro systems. The proposed solution utilizes a BS-BS
CLI orthonormal projector sub-space to near-optimally suppress the
critical BS-BS CLI \textit{on-the-fly}. Compared to the state-of-the-art
dynamic-TDD proposals from industry and academia, the proposed algorithm
offers a significant improvement of the URLLC outage latency performance
and the ergodic capacity accordingly, while greatly minimizing the
control signaling overhead space to $\textnormal{B-bit}$. 

The main insights brought by this paper are as follows: (a) achieving
the URLLC outage targets in dynamic TDD systems are highly challenged
because of the switching delay among the DL and UL transmission opportunities,
and the resultant CLI, (b) the 5G new radio introduces a flexible
slot format design, which in turn minimizes the DL/UL switching delay
to less than a single millisecond, (c) however, within macro deployments,
the BS-BS CLI dominates the URLLC outage performance due to the higher
power DL interfering transmissions, (d) thus, inter-cell CLI coordination
techniques become vital in order to reap the benefits the flexible
TDD systems, and (e) proposed solution demonstrates a near-optimal
BS-BS CLI suppression capability while preserving the transmission
flexibility of the dynamic TDD technology, and with a limited signaling
overhead size.

\section{Acknowledgments}

This work is partly funded by the Innovation Fund Denmark \textendash{}
File: 7038-00009B.

\end{document}